\documentclass[11pt,a4paper]{article}
\usepackage{amsfonts,amsthm,amssymb}
\usepackage[pdftex]{graphicx}
\usepackage{color}
\usepackage[fleqn]{amsmath}
\newcommand{\be}{\begin{eqnarray}}
\newcommand{\ee}{\end{eqnarray}}
\newcommand{\bez}{\begin{eqnarray*}}
\newcommand{\eez}{\end{eqnarray*}}

\theoremstyle{plain}
\newtheorem{theorem}{Theorem}[section]

\theoremstyle{definition}

\newtheorem{remark}[theorem]{Remark}
\newtheorem{example}[theorem]{Example}

\numberwithin{equation}{section}
\numberwithin{theorem}{section}

\begin{document}

\title{\bf \Large Matrix KP: tropical limit, Yang-Baxter and pentagon maps}

\author{
{ \sc{Aristophanes Dimakis}$^a$ and \sc{Folkert M\"uller-Hoissen}$^b$ } 
 \vspace{.3cm} \\
 $^a$ {\small Department of Financial and Management Engineering,} \\
 {\small University of the Aegean, Chios, Greece, e-mail: dimakis@aegean.gr} \\
 $^b$ {\small Max-Planck-Institute for Dynamics and Self-Organization,} \\
        {\small G\"ottingen, Germany, e-mail: folkert.mueller-hoissen@ds.mpg.de} 
}

\date{ }

\maketitle

\begin{abstract}
In the tropical limit of matrix KP-II solitons, their support at fixed time is a planar graph with 
``polarizations'' attached to its linear parts. In this work we explore a subclass of soliton solutions 
whose tropical limit graph has the form of a rooted and generically binary tree, as well as solutions with 
a limit graph consisting of two relatively inverted such rooted tree graphs. The distribution of 
polarizations over the constituting lines of the graph is fully determined by a parameter-dependent 
binary operation and a (in general nonlinear) Yang-Baxter map, which in the vector KP case becomes linear, 
hence is given by an R-matrix. The parameter-dependence of the binary operation leads to a solution 
of the pentagon equation, which exhibits a certain relation with the Rogers dilogarithm via a 
solution of the hexagon equation, the next member in the family of polygon equations.  
A generalization of the R-matrix, obtained in the vector KP case, is found to also solve a pentagon equation. 
A corresponding \emph{local} version of the latter then leads to a new solution of the hexagon equation.  
\end{abstract}

\begin{center}
                  \hspace{-3.1cm} tropical limit, binary tree, dilogarithm.
\end{center}

\section{Introduction}
\label{sec:intro}
In \cite{DMH17} we explored the tropical limit (also see \cite{DMH11KPT,DMH12KPBT,DMH14KdV}) 
of a class of line soliton solutions of the matrix KP-II equation
\be
     ( \, 4 \, u_t - u_{xxx} - 3 \, (u K u)_x \, )_x - 3 \, u_{yy}  
     + 3 \, \Big( u K \int u_y \, dx - \int u_y \, dx \, K u \Big)_x = 0 \, ,
                    \label{KmatrixKP}
\ee
where $K$ is a constant $n \times m$ matrix and $u$ an $m \times n$ matrix, depending on independent variables $x,y,t$. 
A subscript indicates a corresponding partial derivative. We refer to the above equation as KP$_K$. 
In this work we address another class of line soliton solutions, having a rooted, generically binary, tree shaped 
support in the $xy$-plane in the tropical limit, at fixed time $t$. More generally, we will also consider solutions having a 
tropical limit graph which is a kind of superposition of two such rooted trees, one of them upside down.  

In Section~\ref{sec:sol} we recall from \cite{DMH17} a binary Darboux transformation for the KP$_K$ equation 
and describe the class of solutions on which we will focus in this work, as well as their tropical limit. 
Section~\ref{sec:maps} reveals an essential structure of these solutions in the tropical limit. 
The distribution of ``polarizations'' (normalized values of the dependent variable $u$) is ruled by 
a parameter-dependent binary operation together with a Yang-Baxter map, which is in general nonlinear. 
The binary operation satisfies a ``localized'' associativity condition, which then generates a pentagon map, 
a solution of the set-theoretical pentagon equation (see \cite{DMH15} and references cited there). 
In case of a vector KP equation ($n=1$), the Yang-Baxter map becomes linear and is given by an 
R-matrix \cite{DMH17}. 

The solution of the pentagon equation obtained in this way exhibits a certain structure that suggests a generalization, 
which is related to a pentagon identity satisfied by the Rogers dilogarithm. The latter pentagon identity determines 
a ``hexagon map'', a set-theoretical solution of the hexagon equation, which is the next member in the family of 
polygon equations \cite{DMH15}, after the pentagon equation.  
Our generalized pentagon map is then recovered as a 
truncation of this hexagon map. This is explained in Section~\ref{subsec:dilog}.
The hexagon equation first appeared in category theory \cite{Kapr+Voev94PSPM,Stre98}, and later in 
the context of Pachner moves of triangulations of four-dimensional manifolds, 
and corresponding invariants (see, e.g., \cite{Kore14,Kash14,Kash15,Kore+Sady17}). 
The aforementioned hexagon map already appeared in \cite{Kash15}. 

Section~\ref{sec:R->pentagon->hexagon} shows that a generalization of the R-matrix, which shows up in the vector 
KP case, also solves a pentagon equation. Localizating the latter then leads to a 
solution of the hexagon equation. 
Section~\ref{sec:conclusions} contains some concluding remarks.

\section{Soliton solutions of the KP$_K$ equation}
\label{sec:sol}
The potential version (pKP$_K$) of the KP$_K$ equation is  
\be
     4 \phi_{xt} - \phi_{xxxx} - 3 \phi_{yy} - 6 (\phi_x K \phi_x)_x + 6 (\phi_x K \phi_y - \phi_y K \phi_x) = 0 \, , 
                    \label{KmatrixpKP}
\ee
 from which we obtain (\ref{KmatrixKP}) via $u = 2 \, \phi_x$. We recall from \cite{DMH17} 
the following binary Darboux transformation.  
Let $\phi_0$ be a solution of (\ref{KmatrixpKP}). Let $\theta$ and $\chi$ be $m \times N$, respectively $N \times n$, 
matrix solutions of the linear equations  
\bez
  && \theta_y = \theta_{xx} + 2 \phi_{0,x} K \theta \, , \quad
     \theta_t = \theta_{xxx} + 3 \phi_{0,x} K \theta_x + \frac{3}{2} (\phi_{0,y} + \phi_{0,xx}) K \theta \, , \\
  && \chi_y = - \chi_{xx} - 2 \chi K \phi_{0,x} \, , \quad
     \chi_t = \chi_{xxx} + 3 \chi_x K \phi_{0,x} - \frac{3}{2} \chi K (\phi_{0,y} - \phi_{0,xx}) \, .
\eez  
The system
\be
 &&  \Omega_x = - \chi K \theta \, , \nonumber \\
 &&  \Omega_y = - \chi K \theta_x + \chi_x K \theta \, , \nonumber \\
 &&  \Omega_t = - \chi K \theta_{xx} + \chi_x K \theta_x - \chi_{xx} K \theta - 3 \chi K \phi_{0,x} K \theta \, ,
         \label{Omega_eqs}
\ee
is then compatible and can be integrated to yield an $N \times N$ matrix solution $\Omega$. 
If $\Omega$ is invertible, 
\bez
    \phi = \phi_0 - \theta \, \Omega^{-1} \chi   
\eez
is a new solution of (\ref{KmatrixpKP}). If the seed solution $\phi_0$ vanishes, solutions of the linear systems 
are given by 
\bez
    \theta = \sum_{a=1}^A \theta_a \, e^{\vartheta(P_a)} \, , \qquad
      \chi = \sum_{i=1}^M e^{-\vartheta(Q_i)} \, \chi_i  \, ,  
\eez  
where $P_a, Q_i$ are constant $N \times N$ matrices, $\theta_a, \chi_i$ are constant $m \times N$, 
respectively $N \times n$ matrices, and
\be
     \vartheta(P) = x \, P + y \, P^2 + t \, P^3 \, .   \label{vartheta}
\ee
If, for all $a,i$, the matrices $P_a$ and $Q_i$ have no common eigenvalue, the Sylvester equations
\bez
     Q_i W_{ia} - W_{ia} P_a = \chi_i K \theta_a   \qquad a=1,\ldots,A, \quad i=1,\ldots,M ,
\eez
have unique $N \times N$ matrix solutions $W_{ia}$, and (\ref{Omega_eqs}) is solved by 
\bez
    \Omega = \Omega_0 + \sum_{a=1}^A \sum_{i=1}^M e^{-\vartheta(Q_i)} \, W_{ia} \, e^{\vartheta(P_a)} \, ,
\eez
with a constant $N \times N$ matrix $\Omega_0$.

\subsection{The class of solutions with $N=1$}
In this work, we will concentrate on the subclass of soliton solutions with $N=1$. In this case, the matrices $P_a$ 
and $Q_i$ consist of a single entry only, for which we write $p_a$, respectively $q_i$. $\theta_a$, $a=1,\ldots,A$, are 
$m$-component column vectors, and $\chi_i$, $b=1,\ldots,M$, are $n$-component row vectors. The Sylvester equation 
is solved by the constants
\bez
      W_{ia} = \frac{\chi_i K \theta_a}{q_i - p_a} \, .
\eez
The above binary Darboux transformation, with vanishing seed and $\Omega_0=0$, then yields the solution
\bez
     \phi = \frac{1}{\tau} \sum_{a=1}^A \sum_{i=1}^M \phi_{ai} \, \tau_{ai} \, ,
\eez
where 
\bez
   &&  \phi_{ai} = (p_a - q_i) \, \frac{\theta_a \, \chi_i}{\chi_i K \theta_a} = \frac{ \theta_a \, \chi_i }{\mu_{ai}}  \, , \\ 
   &&  \tau = - \Omega = \sum_{a=1}^A \sum_{i=1}^M \tau_{ai} \, , \qquad \tau_{ai} = \mu_{ai} \, e^{\vartheta_{ai}} \, , \\
   &&  \mu_{ai} = \frac{\chi_i K \theta_a}{p_a - q_i}  \, , \qquad 
       \vartheta_{ai} = \vartheta(p_a) - \vartheta(q_i) \, .
\eez
This leads to
\bez
     u &=& \frac{2}{\tau^2} \sum_{a,b=1}^A \sum_{i,j=1}^M \Big( 1 - \frac{p_a-q_i}{p_b-q_j} \Big) \, (\chi_j K \theta_b) \, 
           \theta_a \, \chi_i \, e^{\vartheta_{ai} + \vartheta_{bj}} \\
       &=& \frac{1}{\tau^2} \sum_{a,b=1}^A \sum_{i,j=1}^M u_{ai,bj} \, \tau_{ai} \, \tau_{bj} \, ,
\eez
where 
\bez
     u_{ai,bj} = \frac{1}{2} (p_a - q_i - p_b + q_j) ( \phi_{ai} - \phi_{bj} ) 
\eez
In the following, we will assume that $\mu_{ai} > 0$ for all $a,i$, which ensures regularity of the solution. 
The tropical limit of the above $\tau$-function is 
\bez
    \tau_{\mathrm{trop}} = \mathrm{max}\{\tau_{ai} \, | \, a=1,\ldots,A, \; i=1,\ldots,M \} \, .  
\eez
Let $\mathcal{U}_{ai}$ be the region of $\mathbb{R}^3$, where $\tau_{ai} \geq \tau_{bj}$ for all $b,j$. 
The tropical limit of $\phi$ in this ``dominating phase region'' is given by $\phi_{ai}$, 
which satisfies $\mathrm{tr}(K \phi_{ai}) = p_a - q_i$. The boundary of two phase regions is determined by 
$\tau_{ai} = \tau_{bj}$. The tropical value of $u$ along this boundary is $u_{ai,bj}$. 
We note that the tropical values of $\phi$ and $u$ do not depend on the independent variables $x,y,t$, and also not on $A,M$. 

\begin{example}
 For $A=2$ and $M=1$, or $A=1$ and $M=2$, the solution describes a single line soliton. For $A=M=2$, we have two crossing line solitons. 
$A=1$ and $M=3$ leads to a Y-shaped graph in the $xy$-plane, a Miles resonance. For $A=3$ and $M=1$ we obtain a turned over 
Y-shaped graph. $A=3$ and $M=3$ yields a superposition of the latter two graphs. $A=3$ and $M=2$ yields a superposition of 
a turned over Y-shaped graph and a line. See Fig.~\ref{fig:tropical_limit_examples}. 
\begin{figure}
\begin{center}
\includegraphics[scale=.3]{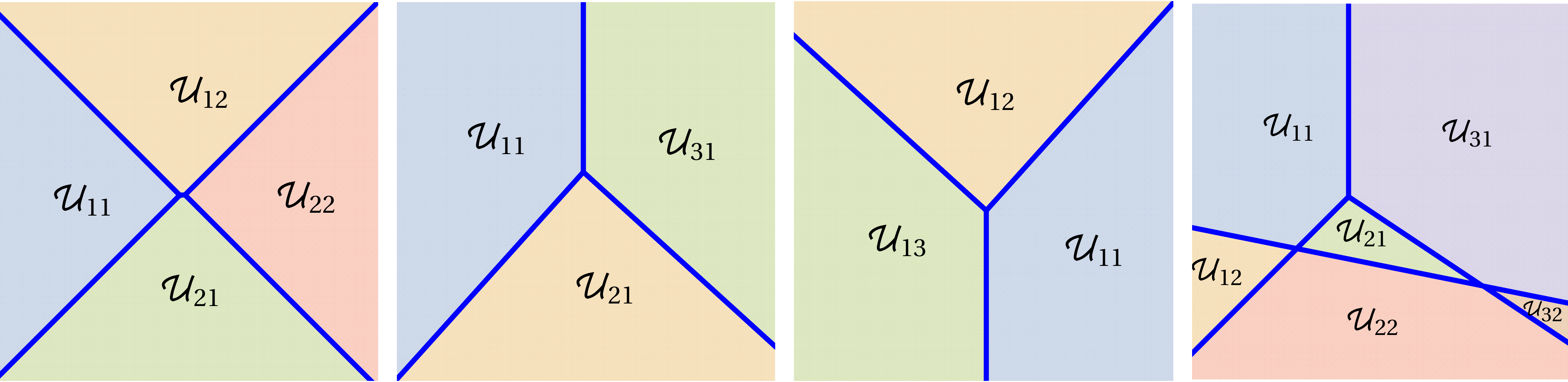} 
\end{center}
\caption{Contour plots of $\tau_{\mathrm{trop}}$ for fixed $t$ in the $xy$-plane, with $(A,M)=(2,2), (3,1), (1,3), (3,2)$, 
respectively. The resulting graph (here with blue lines) divides the $xy$-plane into dominating phase regions $\mathcal{U}_{ai}$. 
\label{fig:tropical_limit_examples}}
\end{figure}
\hfill $\square$
\end{example}

At a coincidence of $L$ phases, i.e., at points in $\mathbb{R}^3$, where $\tau_{a_1,i_1} = \ldots = \tau_{a_L,i_L} > \tau_{b,j}$ 
for all remaining $(b,j)$, the tropical value of $u$ is 
\bez
     u_{a_1,i_1,\ldots,a_L,i_L} = \frac{4}{L^2} \sum_{1 \leq r<s \leq L} u_{a_r i_r, a_s i_s} \, .
\eez

Instead of $u_{ai,bj}$, we will rather consider the modified values
\bez
     \hat{u}_{ai,bj} = \frac{\phi_{ai} - \phi_{bj}}{p_a - q_i - p_b + q_j} \, ,
\eez
which are normalized in the sense that $\mathrm{tr}(K \hat{u}_{ai,bj}) = 1$. For $a=b$ or $i=j$, these are 
rank one projections. The normalized values satisfy the identities
\bez
    ( p_{ai}-p_{bj} ) \, \hat{u}_{ai,bj} + ( p_{bj}-p_{ck} ) \, \hat{u}_{bj,ck} + ( p_{ck}-p_{ai} ) \, \hat{u}_{ck,ai} = 0   \, ,
    \qquad
    p_{ai} = p_a - q_i \, ,
\eez
around (but not at) coincidence points of three dominating phase regions (i.e., points where three lines meet in the $xy$-plane, at some $t$), 
which are determined by $\tau_{ai} = \tau_{bj} = \tau_{ck}$. For $i=j=k$, this reads
\be
    \hat{u}_{ai,ci} 
  = \frac{p_a-p_b}{p_a-p_c} \, \hat{u}_{ai,bi} + \frac{p_b-p_c}{p_a-p_c} \, \hat{u}_{bi,ci}  
  = \frac{p_a-p_b}{p_a-p_c} \, \hat{u}_{ai,bi} + \Big( 1 - \frac{p_a-p_b}{p_a-p_c} \Big) \, \hat{u}_{bi,ci} 
    \; \quad i=1,\ldots,M .  \quad      \label{hatu_identity}
\ee

\section{Maps ruling the distribution of polarizations on the tropical limit graphs}
\label{sec:maps}
\begin{figure}
\begin{center}
\includegraphics[scale=.3]{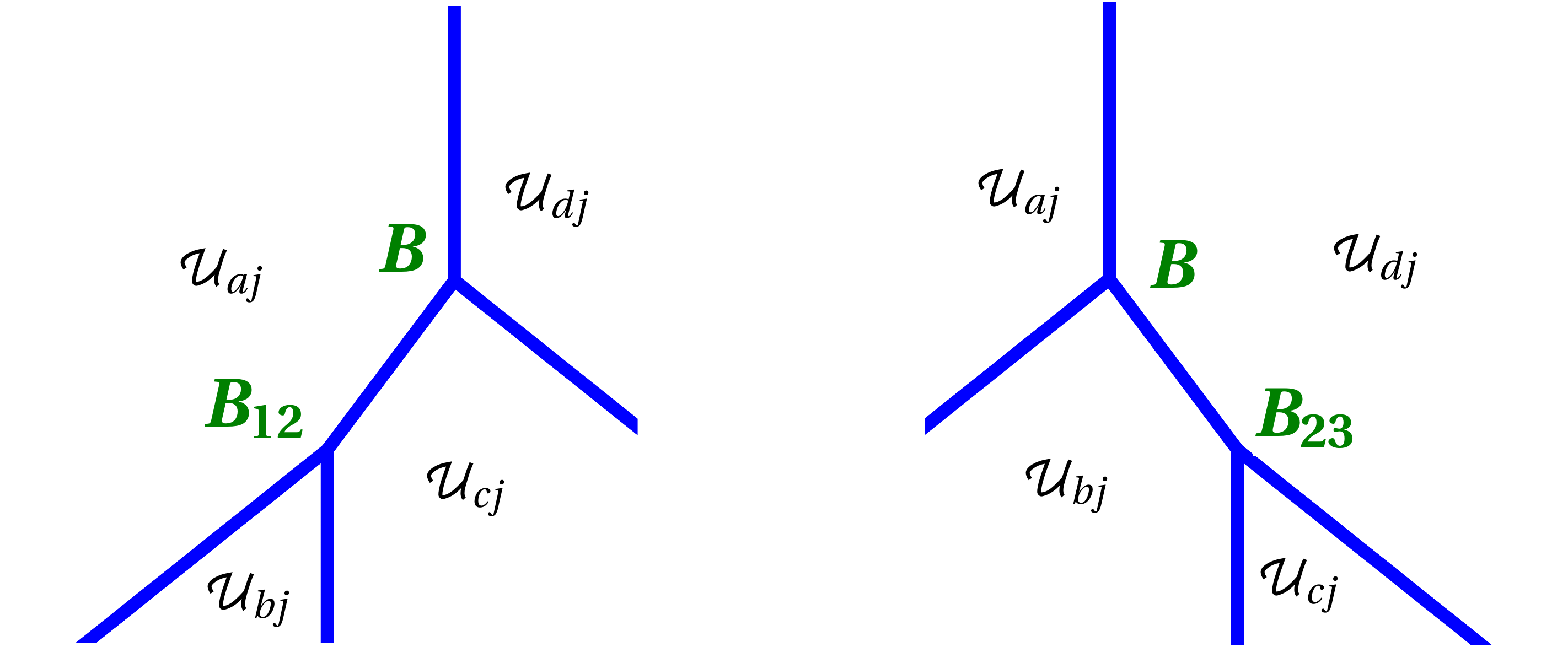} 
\end{center}
\caption{Tropical limit graphs in the $xy$-plane, at times $t \ll 0$ (left) and $t \gg 0$ (right), of a solution with $A=4$ (and arbitrary $M$). 
They provide us with two different, but equivalent ways to map three incoming ($y \ll 0$)  polarizations to a single outgoing ($y \gg 0$) 
polarization. This means that the binary operation $\boldsymbol{B}$ satisfies the tetragon equation.
\label{fig:4solitons_tetragon_eq}}
\end{figure}

\begin{figure}
\begin{center}
\includegraphics[scale=.18]{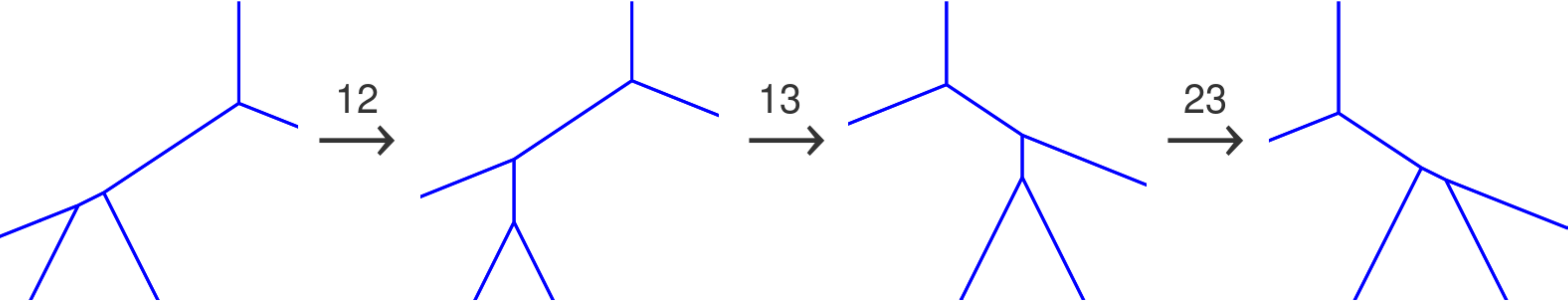} \\
\includegraphics[scale=.15]{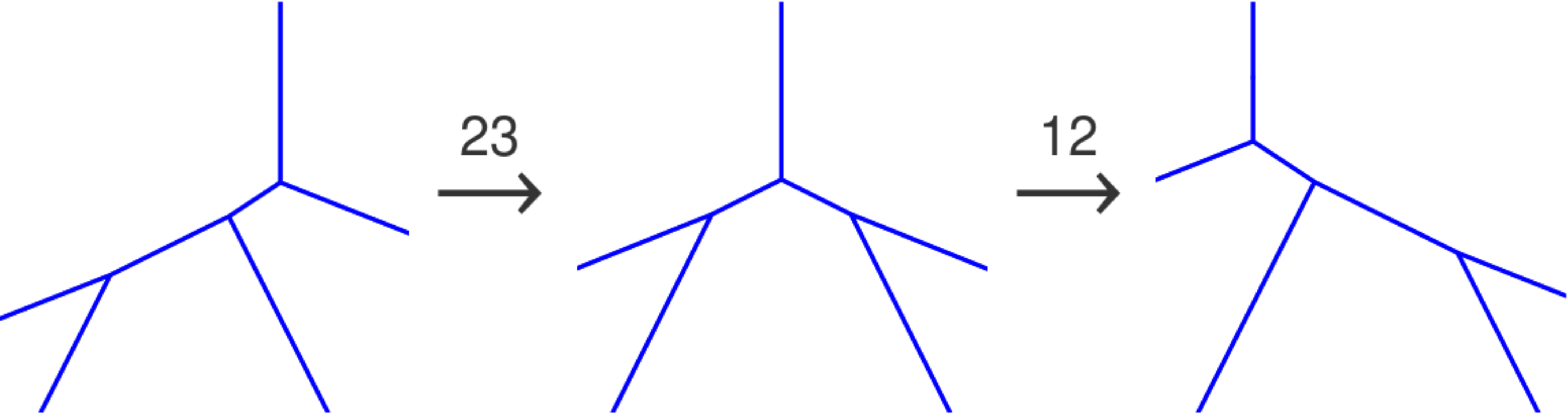}
\end{center}
\caption{Two chains of tropical limit graphs in the $xy$-plane, at consecutive times, of a solution with $A=5$. 
The first chain appears for a negative value of the next KP hierarchy variable $s$, the second chain for a positive value.
Each graph corresponds to a composition of three maps $\boldsymbol{B}$, acting upwards at a vertex of a graph and 
with certain parameters. The two different chains originate from the fact that the associativity condition 
(see Fig.~\ref{fig:4solitons_tetragon_eq}) can be applied in different ways. Each step of a chain, i.e., 
each application of the associativity relation, is accompanied by a map $\mathcal{T}$ of the parameters. 
The binary operation $\boldsymbol{B}$ does not depend on the variables $x,y,t,s$. Therefore the compositions of 
maps $\mathcal{T}$, associated with each chain of graphs, are equivalent, and this imposes the pentagon equation
on $\mathcal{T}$. Also see Example~\ref{ex:5soliton}. 
\label{fig:pentagon}}
\end{figure}

(\ref{hatu_identity}) defines a binary operation
\bez
    \boldsymbol{B}(\lambda) : \; V \times V \longrightarrow V \, , \qquad
    (\xi,\eta) \mapsto  \left( \begin{array}{cc} \xi & \eta \end{array} \right) 
         \left( \begin{array}{c} \lambda \\ 1-\lambda \end{array} \right) 
       = \lambda \, \xi + (1 - \lambda) \, \eta \, ,
\eez
where $V$ is the vector space in which the variables $\hat{u}_{ai,bj}$ take their values. 
In terms of this map, the above identity takes the form
\bez
     (\hat{u}_{ai,bi}, \hat{u}_{bi,ci}) \, \boldsymbol{B}\Big(\frac{p_a-p_b}{p_a-p_c}\Big) = \hat{u}_{ai,ci} \, .
\eez
The binary operation satisfies the \emph{local tetragon equation} (cf. \cite{DMH15}), 
\bez
  \boldsymbol{B}_{\boldsymbol{12}}\Big(\frac{p_a-p_b}{p_a-p_c}\Big) \circ \boldsymbol{B}\Big(\frac{p_a-p_c}{p_a-p_d}\Big) 
  = \boldsymbol{B}_{\boldsymbol{23}}\Big(\frac{p_b-p_c}{p_b-p_d}\Big) \circ \boldsymbol{B}\Big(\frac{p_a-p_b}{p_a-p_d}\Big)  \, ,
\eez  
assuming the denominators to be non-zero. Here boldface indices indicate the positions on which the map $\boldsymbol{B}$ acts, 
from the right, on a threefold direct sum. 
This equation is a parameter-dependent associativity condition, see Fig.~\ref{fig:4solitons_tetragon_eq}. 
As a consequence, the (twisted) map
\be
     \mathcal{T}\Big( \frac{p_a - p_b}{p_a - p_c} , \frac{p_a - p_c}{p_a - p_d} \Big) 
     = \Big( \frac{p_a - p_b}{p_a - p_d} , \frac{p_b - p_c}{p_b - p_d} \Big)   \label{T_map}
\ee
then satisfies the \emph{pentagon equation} (see \cite{DMH15} and references therein)
\bez
     \mathcal{T}_{\boldsymbol{23}} \circ \mathcal{T}_{\boldsymbol{13}} \circ \mathcal{T}_{\boldsymbol{12}}
     = \mathcal{T}_{\boldsymbol{12}} \circ \mathcal{T}_{\boldsymbol{23}} \, ,
\eez 
see Fig.~\ref{fig:pentagon}.

\begin{remark}
For $a=b=c$, (\ref{hatu_identity}) becomes
\bez
  \hat{u}_{ai,ak} = \frac{q_i - q_j}{q_i - q_k} \, \hat{u}_{ai,aj} + \frac{q_j - q_k}{q_i - q_k} \, \hat{u}_{aj,ak} 
                  = \frac{q_i - q_j}{q_i - q_k} \, \hat{u}_{ai,aj} + \Big( 1 - \frac{q_i - q_j}{q_i - q_k} \Big) \, \hat{u}_{aj,ak} \, .
\eez
This determines a similar binary operation as the one we met above, but this one acts along tropical limit graphs in the opposite 
(i.e., negative $y$-) direction. \hfill $\square$
\end{remark}

\begin{figure}
\begin{center}
\includegraphics[scale=.3]{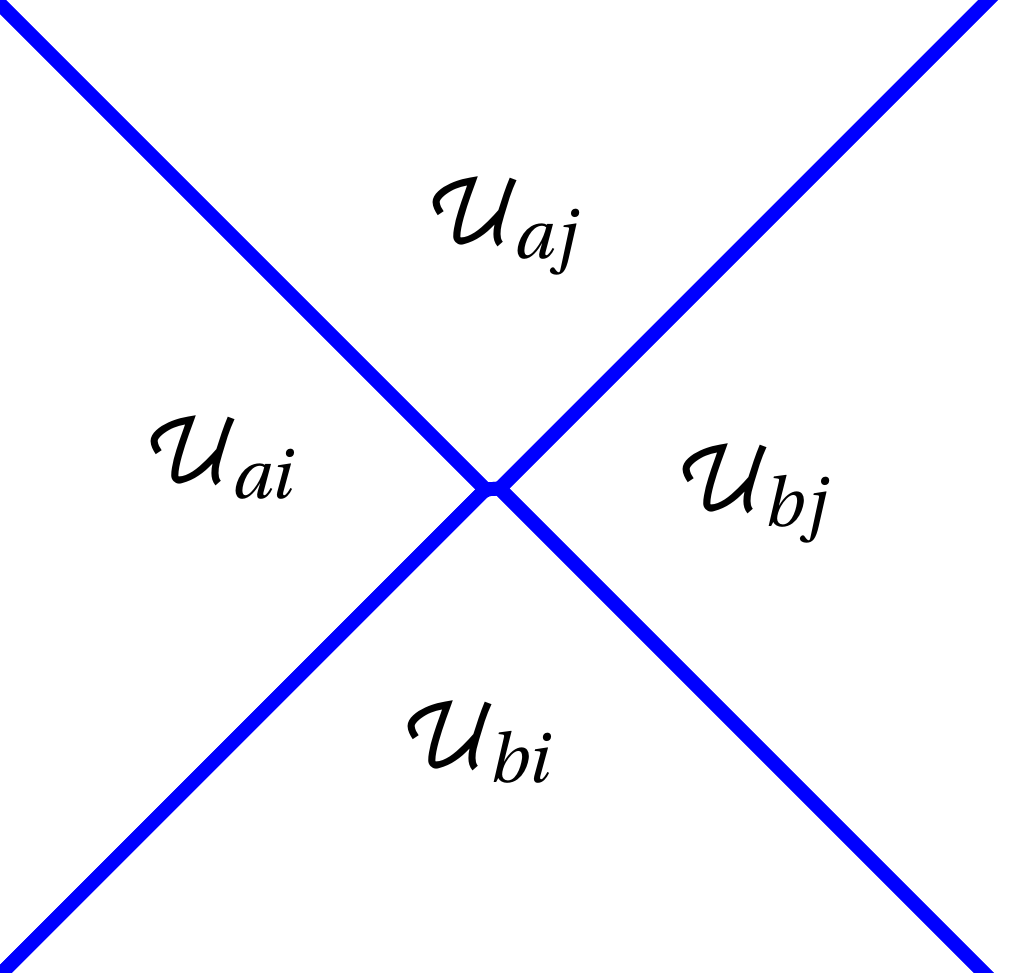} 
\end{center}
\caption{The graph shows the generic situation around a ``crossing'', up to reflections (induced by permutations of $a,b$ or $i,j$). 
It involves a soliton line associated with parameters $p_a,p_b$, and one associated with parameters $q_i,q_j$. 
\label{fig:crossing}}
\end{figure}

If $M=1$ or $A=1$, the tropical limit graph of the soliton solution is a rooted (generically) binary tree. In the first case 
the root is at the top in the $xy$-plane, in the second case it is at the bottom.
If $A,M >1$, the graph is a kind of superposition of two graphs from the latter two classes. In this case crossings appear. 
Their structure is sketched in Fig.~\ref{fig:crossing}.
The (normalized) tropical values of the KP variable, i.e., the polarizations, below and above the crossing are related by a (in general) 
nonlinear Yang-Baxter map $\mathcal{R}(p_a,q_i;p_b,q_j)$, determined by
\be
   && \hat{u}_{aj,bj} = \alpha_{abij}^{-1} \Big(1_m - \frac{q_i - q_j}{p_a - q_j} \, \hat{u}_{bi,bj} \, K \Big) \, \hat{u}_{ai,bi} \, 
                      \Big(1_n - \frac{q_j - q_i}{p_b - q_i} \, K \, \hat{u}_{bi,bj} \Big) \, , \nonumber \\
   && \hat{u}_{ai,aj} = \alpha_{abij}^{-1} \Big(1_m - \frac{p_b - p_a}{p_b - q_j} \, \hat{u}_{ai,bi} \, K \Big) \, \hat{u}_{bi,bj} \, 
                      \Big(1_n - \frac{p_a - p_b}{p_a - q_j} \, K \, \hat{u}_{ai,bi} \Big) \, ,      \label{YB-map}              
\ee
where $1_m$ stands for the $m \times m$ identity matrix, and 
\bez
    \alpha_{abij} = 1 - \frac{(p_a-p_b)(q_j-q_i)}{(p_a-q_j)(p_b-q_i)} \, \mathrm{tr}( K \, \hat{u}_{ai,bi} \, K \, \hat{u}_{bi,bj} ) \, ,
\eez
cf. \cite{DMH17}. The values of $\hat{u}$ in the middle of the right hand sides of the two equations (\ref{YB-map}) are 
the input data of the Yang-Baxter map. This map is invertible. Since (\ref{YB-map}) is still valid if we permute $a$ and $b$, 
or $i$ and $j$, the inverse is obviously obtained by applying both permutations. 
In our examples and figures, it will be convenient to regard the action of the Yang-Baxter map as a process in $y$-direction. 
For a crossing in any concrete example, Fig.~\ref{fig:crossing} (as a graph in the $xy$-plane) is only true for special values 
of $a,b$ and $i,j$, of course. Exchanging $a$ and $b$, respectively $i$ and $j$, would then mean regarding the Yang-Baxter 
map as acting in a different direction in the $xy$-plane. 
 
Whereas the Yang-Baxter equation is actually realized in the case of ``pure solitons'', see \cite{DMH17}, 
such an explicit realization does not exist in the class of solitons considered in this work.

\subsection{The vector KP case}
\label{subsec:vector}
In the vector KP case, i.e., $n=1$, the above Yang-Baxter map becomes linear. Writing $\hat{v}$ instead of $\hat{u}$ in this case, 
we find
\bez
    \left( \begin{array}{cc} \hat{v}_{ai,aj} & \hat{v}_{aj,bj} \end{array} \right) 
    = \left( \begin{array}{cc} \hat{v}_{ai,bi} & \hat{v}_{bi,bj} \end{array} \right) 
      \boldsymbol{R}\Big(\frac{p_a - p_b}{p_a - q_j},\frac{q_j - q_i}{p_a - q_i}  \Big) \, ,
\eez
where
\be
   \boldsymbol{R}(\lambda,\mu) = \left( \begin{array}{cc}
                   \lambda & 1-\mu  \\
                   1-\lambda & \mu \end{array} \right) \, .  \label{R-matrix}
\ee
The Yang-Baxter equation then reads
\bez
  && \boldsymbol{R}_{\boldsymbol{12}}\Big(\frac{p_a - p_b}{p_a - q_j},\frac{q_i - q_j}{p_a - q_j}\Big) \, 
     \boldsymbol{R}_{\boldsymbol{13}}\Big(\frac{p_a - p_c}{p_a - q_k},\frac{q_i - q_k}{p_a - q_k}\Big) \, 
     \boldsymbol{R}_{\boldsymbol{23}}\Big(\frac{p_b - p_c}{p_b - q_k},\frac{q_j - q_k}{p_b - q_k}\Big)  \\
  && \hspace{1cm}  = \boldsymbol{R}_{\boldsymbol{23}}\Big(\frac{p_b - p_c}{p_b - q_k},\frac{q_j - q_k}{p_b - q_k}\Big) \, 
     \boldsymbol{R}_{\boldsymbol{13}}\Big(\frac{p_a - p_c}{p_a - q_k},\frac{q_i - q_k}{p_a - q_k}\Big) \, 
     \boldsymbol{R}_{\boldsymbol{12}}\Big(\frac{p_a - p_b}{p_a - q_j},\frac{q_i - q_j}{p_a - q_j}\Big) \, .
\eez
$\boldsymbol{R}$ and the binary operation $\boldsymbol{B}$ satisfy a consistency condition, which is
\bez
 &&  \boldsymbol{B}_{\boldsymbol{12}}\Big(\frac{p_a - p_b}{p_a - p_c}\Big) \, 
     \boldsymbol{R}\Big(\frac{p_a - p_c}{p_a - q_k},\frac{q_i - q_j}{p_a - q_k}\Big) \\
 && \hspace{1cm}  = \boldsymbol{R}_{\boldsymbol{23}}\Big(\frac{p_b - p_c}{p_b - q_j},\frac{q_i - q_j}{p_b - q_j}\Big) \, 
     \boldsymbol{R}_{\boldsymbol{12}}\Big(\frac{p_a - p_b}{p_a - q_j},\frac{q_i - q_j}{p_a - q_j}\Big) \, 
     \boldsymbol{B}_{\boldsymbol{23}}\Big(\frac{p_a - p_b}{p_a - p_c}\Big) \, . 
\eez

\begin{example}
Fig.~\ref{fig:4soliton_plots} shows plots of $K u$ for a soliton solution with $A=4$ and $M=1$. 
Here we chose $K=(1,1,1,1)$, $\theta_1 = - e_1, \theta_2 = - e_2, \theta_3 = e_3, \theta_4 = e_4$, 
where $e_i$ is the unit vector in $i$-direction, and $p_1=-3/4,p_2=-1/4,p_3=1/4,p_4=3/4$. 
Left and right plot corresponds, respectively, to left and right graph in Fig.~\ref{fig:4solitons_tetragon_eq}.
\begin{figure}
\begin{center}
\includegraphics[scale=.6]{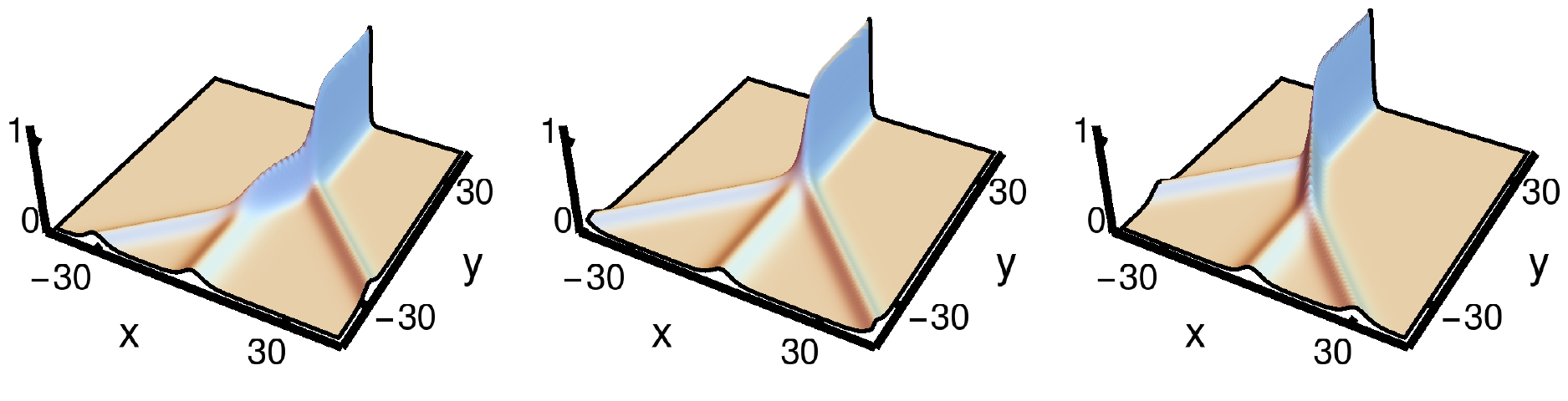} 
\end{center}
\caption{Plots of the scalar $K u$ for a 4-soliton ($A=4$, $M=1$) solution of the $m=4$ vector KP equation at 
times $t=-20,0,20$. 
\label{fig:4soliton_plots}}
\end{figure}
\hfill $\square$
\end{example}

\begin{example}
\label{ex:5soliton}
The time evolution of a rooted tree-shaped KP soliton solution determines a sequence 
of tropical limit graphs, which are rooted binary trees, connected by right rotation in trees  
\cite{DMH11KPT,DMH12KPBT}. At which vertex such a right rotation takes place next, depends on the 
values of higher KP hierarchy evolution variables \cite{DMH11KPT,DMH12KPBT}. In case of a 5-soliton 
solution, we only need the next higher KP hierarchy variable. Correspondingly, we replace (\ref{vartheta}) by 
\bez
   \vartheta(P) = x \, P + y \, P^2 + t \, P^3 + s \, P^4 \, .
\eez
This means that we write explicitly the dependence on the next KP hierarchy evolution variable $s$, which so far 
was hidden in parameters of the solution family. 
Then there are only two structurally different evolutions, described as chains of rooted binary trees, 
depending on whether $s$ is negative or 
positive. These are the chains in Fig.~\ref{fig:pentagon}. 
Since the distribution of tropical limit vectors (polarizations) over a tropical limit graph does not depend on the 
independent variables $x,y,t,s$, the ``outgoing'' (i.e., for $y\gg0$, on the root) polarization computed from four incoming 
($y\ll0$) polarizations yields the same result in case of the first graph of the first chain and the first graph of 
the second chain. The same holds for the last graph of the first chain and the last graph of the second chain.   
This then implies (also see, e.g., \cite{DMH15}) that the map $\mathcal{T}$ satisfies the pentagon equation, as already 
explained in the caption of Fig.~\ref{fig:pentagon}. The latter actually shows the two chains of tropical limit graphs for 
$K=(1,1,1,1,1)$ and the following choice of parameters,
\bez
 && \theta_1 = \left(\begin{array}{c} -1 \\ 0 \\ 0 \\ 0 \\ 0 \end{array} \right) , \;
    \theta_2 = \left(\begin{array}{c} 0 \\ -1 \\ 0 \\ 0 \\ 0 \end{array} \right) , \;
    \theta_3 = \left(\begin{array}{c} 0 \\ 0 \\ -1 \\ 0 \\ 0 \end{array} \right) , \;
    \theta_4 = \left(\begin{array}{c} 0 \\ 0 \\ 0 \\ -1 \\ 0 \end{array} \right) , \;
    \theta_5 = \left(\begin{array}{c} 0 \\ 0 \\ 0 \\ 0 \\ 1 \end{array} \right) , \\
 &&  p_1 = -2 \, , \quad 
     p_2 = - \frac{1}{2} \, , \quad 
     p_3 = 0 \, , \quad
     p_4 = \frac{1}{2} \, , \quad
     p_5 = 2 \, , \quad 
     q_1 = 1 \, , \quad
     \eta_1 = 1 \, .  
\eez
The graphs of the first chain in Fig.~\ref{fig:pentagon} are obtained at times $t=-60,-20,20,80$ and with $s=-20$. 
Those of the second chain are obtained with $t=-30,0.2,30$ and $s=20$. 
\hfill $\square$
\end{example}

\subsection{Relation with the pentagon identity of the dilogarithm}  
\label{subsec:dilog}
Setting
\bez
    X = \frac{p_a - p_b}{p_a - p_c} \, , \quad
    Y = \frac{p_a - p_c}{p_a - p_d} \, , \quad
    X' = \frac{p_a - p_b}{p_a - p_d} \, , \quad
    Y' = \frac{p_b - p_c}{p_b - p_d} \, ,
\eez
we have 
\bez
    X' = X Y \, , \qquad  
    Y' = \frac{Y-XY}{1-XY} \, ,
\eez
so that (\ref{T_map}) reads
\bez
    \mathcal{T}(X,Y) = \Big( X Y \, , \, \frac{Y-XY}{1-X Y} \Big) \, .
\eez
This pentagon map already appeared in the context of the dilogarithm \cite{Kash+Serg98,Kash99TMP}. 
The pentagon identity for the Rogers dilogarithm \cite{Roge07} reads 
\bez
  L(X) + L(Y) = L\Big(\frac{Y-XY)}{1 - XY}\Big) + L(X Y) + L\Big(\frac{X-XY}{1-XY}\Big) \, .
\eez
It implies that the map defined by \cite{Kash15}
\bez
    S(X,Y) = \Big( \frac{Y-XY}{1-XY},XY,\frac{X-XY}{1-XY} \Big)
\eez
solves the hexagon equation (see \cite{DMH15}, for example)
\be
    S_{\boldsymbol{12}} \circ S_{\boldsymbol{23}} \circ \mathcal{P}_{\boldsymbol{34}} \circ S_{\boldsymbol{12}} 
  = \mathcal{P}_{\boldsymbol{34}} \circ S_{\boldsymbol{45}} \circ S_{\boldsymbol{23}} \circ \mathcal{P}_{\boldsymbol{12}} 
    \circ S_{\boldsymbol{23}} \, ,   \label{hexagon_eq}
\ee
where $\mathcal{P}(X,Y)=(Y,X)$. The composition $\hat{\mathcal{T}} = \mathcal{P} \circ \mathcal{T}$, with the above 
map $\mathcal{T}$, is obtained from $S$ by disregarding the last component of its range. It satisfies 
the pentagon equation in the form 
$\hat{\mathcal{T}}_{\boldsymbol{12}} \circ \hat{\mathcal{T}}_{\boldsymbol{23}} \circ \hat{\mathcal{T}}_{\boldsymbol{12}} 
 = \hat{\mathcal{T}}_{\boldsymbol{23}} \circ \mathcal{P}_{\boldsymbol{12}} \circ \hat{\mathcal{T}}_{\boldsymbol{23}}$. 
The above hexagon equation appeared in a slightly different, but equivalent form in \cite{Kash15} (see (3.8) therein). 
Its graphical version (3.7) in \cite{Kash15} coincides with that displayed on page 189 of \cite{Stre98}. 

\begin{remark}
We note that there is a generalization of the above map $\mathcal{T}$ with $X,Y$ from any non-commutative 
associative algebra, which also satisfies the pentagon equation:
\bez
    \mathcal{T}(X,Y) = \left( X Y \, , \, (1-YX)^{-1} \, Y \, (1-X) \right) \, .
\eez
\vspace{-.8cm}

\hfill $\square$
\end{remark}

\section{From the vector KP R-matrix to solutions of the pentagon and hexagon equation}
\label{sec:R->pentagon->hexagon}
We observe that, with a different choice of parameters, the R-matrix obtained in Section~\ref{subsec:vector} also solves 
the pentagon equation,
\bez
 &&  \boldsymbol{R}_{\boldsymbol{23}}\Big(\frac{p_2 - p_5}{p_2 - p_4},\frac{p_3 - p_4}{p_2 - p_4}\Big) \, 
     \boldsymbol{R}_{\boldsymbol{13}}\Big(\frac{p_1 - p_5}{p_1 - p_4},\frac{p_2 - p_4}{p_1 - p_4}\Big) \, 
     \boldsymbol{R}_{\boldsymbol{12}}\Big(\frac{p_1 - p_4}{p_1 - p_3},\frac{p_2 - p_3}{p_1 - p_3}\Big)  \\
 && \hspace{1cm}  = \boldsymbol{R}_{\boldsymbol{12}}\Big(\frac{p_1 - p_5}{p_1 - p_3},\frac{p_2 - p_3}{p_1 - p_3}\Big) \, 
     \boldsymbol{R}_{\boldsymbol{23}}\Big(\frac{p_1 - p_5}{p_1 - p_4},\frac{p_3 - p_4}{p_1 - p_4}\Big) \, .
\eez
This means that 
\bez
   T_{ijkl} := \boldsymbol{R}\Big(\frac{p_i - p_l}{p_i - p_k},\frac{p_j - p_k}{p_i - p_k}\Big)
  = \left(
 \begin{array}{cc}
 \frac{p_i-p_l}{p_i-p_k} & \frac{p_i-p_j}{p_i-p_k} \\[.5em]
  \frac{p_l-p_k}{p_i-p_k} & \frac{p_j-p_k}{p_i-p_k} 
 \end{array}
    \right)  
\eez
satisfies the pentagon equation in the form
\bez
   T_{2345,\boldsymbol{23}} \, T_{1245,\boldsymbol{13}} \, T_{1234,\boldsymbol{12}} 
    = T_{1235,\boldsymbol{12}} \, T_{1345,\boldsymbol{23}} \, .
\eez
According to its origin (\ref{R-matrix}), the matrix $T_{ijkl}$ has the structure 
\bez
     T(\lambda,\mu) = \left( \begin{array}{cc} \lambda & 1-\mu \\ 
                                       1-\lambda & \mu
                              \end{array} \right) \, .
\eez
The \emph{local} pentagon equation
\bez
      T_{\boldsymbol{23}}(X_3,Y_3) \, T_{\boldsymbol{13}}(X_2,Y_2) \, T_{\boldsymbol{12}}(X_1,Y_1) 
    = T_{\boldsymbol{12}}(x_1,y_1) \, T_{\boldsymbol{23}}(x_2,y_2) 
\eez
then determines a map $\mathcal{Q}$ via $\mathcal{Q}(x_1,y_1;x_2,y_2) = (X_3,Y_3;X_2,Y_2;X_1,Y_1)$, which is
\bez
  && Q(x_1,y_1;x_2,y_2) \\
  &=& \Big( \frac{x_2 \, (x_1+y_1-1)}{x_1-x_2+x_2 y_1},\frac{y_2}{y_1+y_2-y_1 y_2};A,y_1+y_2-y_1 y_2;
                         \frac{x_1}{A},\frac{y_1 \, (x_2 y_1+x_1-x_2) (1-x_2-y_2)}{A \, (y_1+y_2-x_1 y_2-x_2 y_1-y_1 y_2)} \Big),
\eez
where
\bez
   A = \frac{x_1 y_1 + x_2 y_2 - x_1 x_2 y_1 - x_1 x_2 y_2 - x_1 y_1 y_2  - x_2 y_1 y_2 + x_1 x_2 y_1 y_2}{y_1+y_2-x_1 y_2-x_2 y_1-y_1 y_2} \, .
\eez
As a consequence (see \cite{DMH15}, for example), this map satisfies the hexagon equation (\ref{hexagon_eq}).

\section{Concluding remarks}
\label{sec:conclusions}
 From our previous work \cite{DMH11KPT,DMH12KPBT} about tree-shaped soliton solutions of the scalar KP 
equation, non-trivial solutions of the pentagon equation were expected to emerge in case of a matrix version 
of the KP equation. We confirmed this in the present work. Moreover, we demonstrated that, for the larger class of 
soliton solutions explored in this work, the distribution of polarizations on the tropical limit graph is ruled 
by a binary operation together with the Yang-Baxter map obtained in \cite{DMH17}, which simplifies to an R-matrix 
in the vector KP case. 

Since the binary operation is parameter-dependent and satisfies a local tetragon equation,
it determines a solution of the pentagon equation, which turned out to be somewhat indirectly related 
to the pentagon identity satisfied by the Rogers dilogarithm, namely via a solution of the hexagon equation induced 
by the latter. 

We also observed that a generalization of the R-matrix obtained in the vector KP case (also see \cite{DMH17}) not only solves 
the Yang-Baxter equation, but also provides us with a solution of the pentagon equation. Its parameter-dependence 
led to an apparently new solution of the hexagon equation.  
\vspace{.3cm}

\noindent
\textbf{Acknowledgments.} F. M.-H. thanks the organizers of the conference ``Physics and Mathematics 
of Nonlinear Phenomena 2017: 50 years of I.S.T.'', where some of the results of this work have been 
reported.

\end{document}